\documentclass[aps,prb,twocolumn,groupedaddress,amsmath,amssymb]{revtex4}
\usepackage{graphicx,mathptm,bbm}
\usepackage{dcolumn}
\usepackage{bm}
\usepackage[colorlinks,linkcolor=red,citecolor=blue,urlcolor=blue]{hyperref}
\usepackage{amsthm,amsbsy}
\usepackage[usenames]{color}

\newcommand{\half}{\mbox{$\textstyle \frac{1}{2}$}}
\newcommand{\ket}[1]{\left | \, #1 \right \rangle}
\newcommand{\kets}[1]{| \, #1 \rangle}
\newcommand{\bra}[1]{\left \langle #1 \, \right |}
\newcommand{\braket}[2]{\left\langle\, #1\,|\,#2\,\right\rangle}

\newcommand{\av}[1]{\langle #1\rangle}

\newcommand{\eqr}[1]{Eq.~(\ref{#1})}

\newcommand{\fir}[1]{Fig.~\ref{#1}}

\definecolor{DarkRed}{rgb}{0.7,0,0}

\begin{document}

\title{Dephasing enhanced transport in non-equilibrium strongly-correlated quantum systems}

\author{J. J. Mendoza-Arenas$^1$, T. Grujic$^1$, D. Jaksch$^{1,2}$ and S. R. Clark$^{2,1}$}
\affiliation{$^1$Clarendon Laboratory, University of Oxford, Parks Road, Oxford OX1 3PU, United Kingdom}
\affiliation{$^2$Centre for Quantum Technologies, National University of Singapore, 3 Science Drive 2, Singapore 117543}
\date{\today}

\begin{abstract}
A key insight from recent studies is that noise, such as dephasing, can improve the efficiency of quantum transport by suppressing coherent single-particle interference effects. However, it is not yet clear whether dephasing can enhance transport in an interacting many-body system. Here we address this question by analysing the transport properties of a boundary driven spinless fermion chain with nearest-neighbour interactions subject to bulk dephasing. The many-body non-equilibrium stationary state is determined using large scale matrix product simulations of the corresponding quantum master equation. We find dephasing enhanced transport only in the strongly interacting regime, where it is shown to induce incoherent transitions bridging the gap between bound dark-states and bands of mobile eigenstates. The generic nature of the transport enhancement is illustrated by a simple toy model, which contains the basic elements required for its emergence. Surprisingly the effect is significant even in the linear response regime of the full system, and it is predicted to exist for any large and finite chain. The response of the system to dephasing also establishes a signature of an underlying non-equilibrium phase transition between regimes of transport degradation and enhancement. The existence of this transition is shown not to depend on the integrability of the model considered. As a result dephasing enhanced transport is expected to persist in more realistic non-equilibrium strongly-correlated systems. 
\end{abstract}


\maketitle

\section{Introduction} \label{intro}
Recently the effects of noise on the efficiency of quantum transport phenomena have been scrutinised intensely by the scientific community. This has been motivated in part by a series of ground-breaking nonlinear spectroscopic experiments on light-harvesting complexes demonstrating surprisingly long-lived quantum coherence during exciton transport, even in a warm and wet environment~\cite{engel07,lee07,collini10}. Yet for purely coherent exciton dynamics in such protein pigment networks, transport is highly suppressed due to destructive interference between different propagation pathways. Instead studies revealed that the remarkably high transport efficiency observed (above 95\%) in fact emerges in combination with local noise, such as dephasing, which disrupts this interference opening up previously inhibited pathways for transmission~\cite{plenio08,mohseni08,caruso09,rebentrost09,chin10,olayacastro08,manzano13}. Transport properties in open systems can thus not only defy the traditional understanding of when quantum effects should play a significant role, but also challenge the notion that couplings to the environment unconditionally degrade performance.

Different effects of dephasing have been studied in networks populated by single particles, in scenarios such as transport through quantum optical systems \cite{caruso11}, heat transport through chains of two-level systems \cite{manzano12}, information transmission \cite{caruso10} and quantum information processing \cite{bermudez12}. Also, interesting phenomena have been seen to emerge from the coexistence of particle-particle interactions and noise, such as glassy dynamics in ordered systems \cite{poletti12a} and interaction impeded decoherence \cite{poletti12b,poletti12c}. However, the interplay between noise and strong correlations induced by interactions in a many-body setting is not yet fully understood. In particular a recent cold-atom experiment showed that the time-of-flight expansion of a strongly-interacting gas was slow in the absence of noise and substantially increased once noise was added~\cite{errico12}. This raises an important question as to when and how dephasing can enhance transport in a strongly interacting system.

Here we answer this question in the affirmative by considering a concrete example composed of spinless fermions with nearest-neighbour interactions hopping through a tight-binding chain, as depicted in \fir{fig1}. This model makes an ideal testbed for several reasons. First, it is equivalent to the well studied $XXZ$ spin-$1/2$ chain~\cite{gobert05,langer09,kessler12}, representing one of the simplest models of strongly-correlated electron systems. Second, the transport properties of such low-dimensional interacting quantum systems remains an important open problem, exhibiting anomalous features such as ballistic spin transport and unusually high thermal conductivity, reported experimentally in so-called spin-chain materials~\cite{sologubenko07,heidrich07,hlubek10,janson10}. In addition to solid state systems like chains of coupled quantum dots~\cite{hanson07} or molecular wires embedded between electrodes~\cite{cuevas10}, understanding this model is directly relevant to ion-trap~\cite{barreiro11}, coupled-cavity array~\cite{kay08}, and cold-atom~\cite{trotzky08,simon11,errico12,ronzheimer13} quantum systems. Of particular importance are recent seminal experiments that revealed contact and bulk resistivity of cold fermionic atoms flowing through a narrow mesoscopic channel between a pair of reservoirs with a population imbalance~\cite{brantut12,stadler12}.

\begin{figure}[t]
\includegraphics{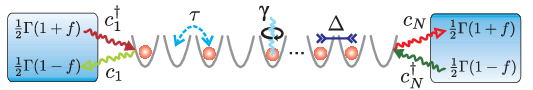}
\caption{Spinless fermions hop with amplitude $\tau$ across a chain, subject to a nearest-neighbour density-density interaction of strength $\Delta$, local dephasing at a rate $\gamma$, and boundary driving that injects/ejects fermions at a rate proportional to $\Gamma$ and driving bias $f$. The driving process induces a forward (upper arrows) and a backward (lower arrows) flow of particles, and the bias $f$ determines the imbalance between both.} \label{fig1}
\end{figure}

Very similar to these cold-atom experiments, we consider a chain attached to two unequal Markovian reservoirs at its boundaries providing continuous incoherent driving, along with local dephasing noise along its extension (see \fir{fig1})~\cite{benenti09a,benenti09b,prosen11a,znidaric11b}. We then compute the current-driving characteristics of its non-equilibrium stationary state (NESS). Being a homogeneous chain with transport from end to end, previously studied single-particle interference effects originating from geometry or disorder are absent~\cite{caruso09,rebentrost09,znidaric10b,kassal12}. Interestingly, we show that dephasing-enhanced transport nonetheless emerges so long as the interactions are strong enough. In this regime and at maximal driving, the NESS forms a cooperative many-body quantum state possessing a long-ranged domain of particles pinned to one boundary strongly suppressing the current, analogous to a Coulomb or Pauli blockade insulator~\cite{benenti09b}. Dephasing induces incoherent transitions out of this bound state establishing a current in a regime that would otherwise be insulating. We illustrate this mechanism in a simple toy model, in which we isolate the essential conditions for the emergence of the effect. Surprisingly, a large current enhancement is predicted to exist in any long and finite chain even for weak driving, where the transport is diffusive in the absence of noise. We also observe that the transport enhancement is a signature of an underlying non-equilibrium phase transition, and demonstrate its generality beyond the integrable system considered.

The paper is organized as follows. In Section \ref{model} we describe the system to be studied. In Section \ref{enhancement} we show the existence of dephasing-assisted transport for strong interactions, which contrasts with the transport degradation at weak interactions. The mechanism behind this non-equilibrium phenomenon is explained in Section \ref{mechanism}. An illustrative toy model containing the basic features for the effect to emerge is described in Section \ref{toy}. The signatures of a non-equilibrium phase transition between the two transport regimes, revealed by the existence of an optimal dephasing rate and the correlations through the system, are presented in Section \ref{phase}. We also show in this Section that this transition remains even if the integrability of the model is broken. Finally, in Section \ref{conclu} we discuss the conclusions of our work. 

\section{Model} \label{model}
We study the $N$ site interacting spinless fermion chain described by the Hamiltonian
\begin{equation} \label{hami}
H =\sum_{j=1}^{N-1}\left[\half \tau(c^\dagger_j c_{j+1}+\textrm{h.c.})+\Delta(n_j - \half)(n_{j+1} - \half)\right],
\end{equation}
where $c^\dagger_j,c_j$ are standard fermionic creation/annihilation operators for site $j$ and $n_j = c^\dagger_jc_j$ is the associated number operator. In addition to the hopping amplitude $\tau$, this Hamiltonian has a nearest-neighbour density-density interaction $n_j n_{j+1}$ with strength $\Delta$. We take $\tau=1$ to set the energy scale. The dynamics of the system is described by a Lindblad quantum master equation~\cite{breuer02} (taking $\hbar = 1$)
\begin{equation} \label{master_eq}
 \frac{d\rho}{dt}=-i[H,\rho] + \mathcal{L}(\rho),
\end{equation} 
where $\rho$ is the density matrix of the chain, and $\mathcal{L}$ is the dissipator describing the coupling to the Markovian reservoirs. In Lindblad form the dissipator is 
\begin{equation}
\mathcal{L}(\rho) = \sum_{k}\biggl(L_{k}\rho L_{k}^{\dagger}-\frac{1}{2}\{L_{k}^{\dagger}L_{k},\rho\}\biggr), 
\end{equation}
where $\{.,.\}$ is the anti-commutator and the sum is over a set of jump operators $L_k$. We consider a dissipator that splits into three parts $\mathcal{L} = \mathcal{L}_L + \mathcal{L}_d + \mathcal{L}_R$. Here $\mathcal{L}_L$ and $\mathcal{L}_R$ describe the coupling to external particle reservoirs at the left and right boundaries, respectively, each with two jump operators 
\begin{equation} \label{jump_driving}
L_{L,R}^+=\sqrt{\Gamma(1\mp f)/2}\,c_{1,N},\quad L^-_{L,R}=\sqrt{\Gamma(1\pm f)/2}\, c^\dagger_{1,N}, 
\end{equation}
where $\Gamma$ is the coupling strength, identical for both reservoirs, and $0\leq f \leq1$ is the driving bias~\cite{benenti09b}. We consider moderate coupling $\Gamma=1$ throughout this paper~\cite{footnote1}. The driving process, depicted in \fir{fig1}, induces two pumping processes, corresponding to forward (left-to-right) and backward (right-to-left) flows, thus forcing the system far-from-equilibrium. This scheme is reminiscent of the well studied classical stochastic exclusion model~\cite{derrida98,golinelli06}. When $f=0$ particles are injected and ejected with equal rates at both boundaries, so the counter-propagating flows cancel each other. This results in the stationary solution $\rho = \mathbbm{1}/2^N$ to \eqr{master_eq}, irrespective of $\Delta$, thus having no net current~\cite{burgarth07}. For $f>0$ the forward flow is favoured over the backward flow, raising the possibility of a genuine NESS possessing a finite current. The remaining contribution $\mathcal{L}_d$ accounts for bulk dephasing in the chain and is described by the jump operators
\begin{equation} \label{dephasing}
L_{j}^d = \sqrt{\gamma}(\mathbbm{1}-2n_j),\qquad 1\leq j\leq N,
\end{equation}
with a uniform dephasing rate $\gamma$.

By directly simulating \eqr{master_eq} and taking the long time limit, the state $\rho(t)$ converges to the time-independent NESS of the system. A solution can be computed efficiently in a controlled way and accounting for significant many-body correlations by applying the time evolving block decimation (TEBD) algorithm \cite{zwolak04,verstraete04} to a matrix product operator description of $\rho(t)$. This highly compact representation enables the efficient evaluation of relevant expectation values and makes accessible much larger system sizes than exact diagonalization or Monte Carlo approaches~\cite{michel08,wu11,popkov12}. Moreover TEBD can be applied effectively over a large parameter range allowing us to examine the properties of the system as a function of $f$ beyond the $f \ll 1$ linear response regime. Our implementation of the numerical method is based on the open source Tensor Network Theory (TNT) library \cite{tntlib}.

The transport properties are analyzed by computing the current crossing site $j$ from the operator
\begin{equation}
J_j = i(c^\dagger_j c_{j+1} - \textrm{h.c.}),\qquad 1\leq j\leq N-1. 
\end{equation}
In the NESS the current $\av{J_j} = \av{J}$ is homogeneous throughout the system.

\begin{figure}[t]
\includegraphics{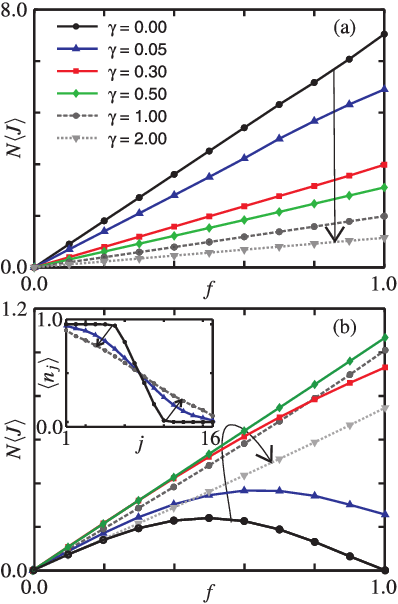}
\caption{(a) The current-driving profiles in the weakly-interacting regime $\Delta=0.5$ for increasing (as indicated by the arrow) dephasing rates $\gamma$ and $N=16$. (b) Identical plot to (a) but in the strongly-interacting regime $\Delta=2$. Inset. The density $\av{n_j}$ of the system when $\Delta = 2$, $f=1$, and $N=16$, corresponding to $\gamma=0,0.05,1.00$.} \label{fig2}
\end{figure}

\section{Dephasing enhanced transport} \label{enhancement}
It is known that in the weakly interacting regime $|\Delta|< 1$, in the absence of dephasing, the system is an ideal ballistic conductor for any driving $f$, with a nearly flat density profile $\av{n_j}$ and a current $\av{J} \propto f$ which is independent of $N$~\cite{prosen09,benenti09b}. The introduction of dephasing has been shown to induce diffusive transport, where the current fulfills the diffusion equation $\av{J}=\kappa\nabla\av{n_j}$, with $\kappa$ the particle conductivity, and $\av{n_j}$ features a constant gradient
\begin{equation}
\nabla \av{n_j} =\frac{\av{n_{N-1}} - \av{n_{2}}}{N-3} = \frac{\Delta n}{N-3},
\end{equation}
where $\Delta n$ is the density difference between opposite ends of the system (after discarding the boundary sites). As a result the current scales with the size of the system as
\begin{equation}
\av{J}\propto\frac{\Delta n}{N-3}\sim \frac{f}{N},
\end{equation}
characteristic of an Ohmic conductor~\cite{znidaric10a,znidaric10b}. In either case the maximum current through the chain occurs at maximal bias $f=1$, where only forward pumping is present, as might be intuitively expected. In the non-interacting limit $\Delta = 0$ it has been proven rigorously that a homogeneous chain cannot exhibit any dephasing enhanced end-to-end transport \cite{znidaric10a,plenio08,kassal12}; this behaviour was also suggested for weakly-interacting systems where $|\Delta|<1$~\cite{znidaric10b}. In \fir{fig2}(a) we report the current-driving profiles for $\Delta = 0.5$ showing that dephasing monotonically degrades the current for any driving, confirming that this behaviour persists even in the presence of weak interactions. Thus, in this work we focus on the strongly interacting regime $|\Delta|> 1$. 

\begin{figure}
\includegraphics{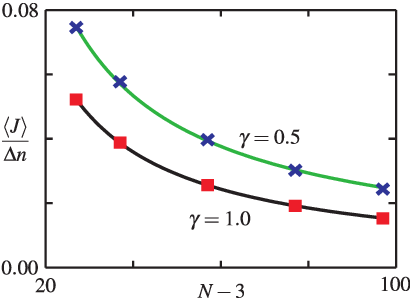}
\caption{The current $\av{J}$ at $f=1$, divided by the resulting density difference $\Delta n$ between the ends of the system (excluding the boundary), is plotted against the system size $N-3$ up to $N=100$ sites. The data is for dephasing rates $\gamma = 0.5$ ($\times$) and $\gamma = 1.0$ ($\blacksquare$). The solid lines show the fitting to $\av{J}/\Delta n = \kappa (N-3)^{-\alpha}$ for each $\gamma$. These yield $\kappa = 1.288$ and $\alpha = 0.863$ for $\gamma = 0.5$, and $\kappa = 1.228$ and $\alpha = 0.958$ for $\gamma = 1.0$.}
\label{fig3} 
\end{figure} 

In the absence of dephasing and for weak driving $f \ll 1$, transport was found to be diffusive when $|\Delta| > 1$~\cite{prosen09}, a controversial finding given that the integrability of the system is conjectured to lead to ballistic transport~\cite{zotos97}. However, for strong driving $f \rightarrow 1$ it was recently discovered~\cite{benenti09b,prosen11b} that the NESS exhibits a particle domain at the left edge of the chain irrespective of the sign of $\Delta$, strongly suppressing the current as $\av{J} \propto \exp(-N)$, characteristic of an insulator. Consequently, the current $\av{J}$ at $\gamma = 0$ exhibits non-linear behaviour with the driving $f$, leading to an effect known as negative differential conductivity (NDC) where increasing the driving eventually decreases the current~\cite{benenti09b,prosen11b}. In \fir{fig2}(b) the $\gamma = 0$ curve shows that this causes a near complete suppression of the current at $f=1$ for $\Delta=2$. In the strongly interacting regime the system therefore presents the intriguing property that more current forward flows at an intermediate bias $f<1$ where some backward pumping is present.

\subsection{Current enhancement}

The main result of the present work is that for $|\Delta|>1$ the presence of a small bulk dephasing can significantly enhance the particle transport. This striking behaviour is illustrated for $\Delta = 2$ in \fir{fig2}(b) where dephasing up to a moderate rate $\gamma \approx 0.5$ is seen to increase the current. This enhancement in the current is shown to occur {\em for any} $f>0$; however it is not uniform in $f$, resulting in the current-driving profile changing with $\gamma$. Specifically, around $\gamma \approx 0.3$ the NDC effect is lost and further increases in $\gamma$ yield a linear profile in $f$. Near $f=1$ dephasing therefore induces not just a quantitative increase in the current, but rather causes a major qualitative change in the behaviour of the system from being insulating at $\gamma = 0$ to yielding the maximal current once $\gamma > 0.3$. In the inset of \fir{fig2}(b) this change at $f=1$ is shown to be coincident with the breakdown of the particle domain at the left boundary into a nearly linear density profile, due to the increase of the dephasing rate. The transport at large driving and dephasing rates therefore resembles that of a diffusive conductor. This result is confirmed by the scaling of the current with the size of the system, shown in \fir{fig3} for $\Delta=2$ and dephasing rates $\gamma=0.5$ and $\gamma=1.0$. The power-law fits for each $\gamma$, which are seen to accurately model the data, indicate sub-diffusive transport for $\gamma = 0.5$, where the current decays slower than $1/(N-3)$, but has approached a diffusive behaviour once $\gamma=1.0$.
\begin{figure}[t]
\includegraphics{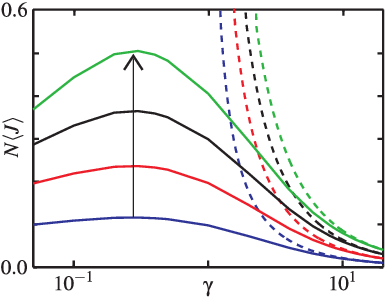}
\caption{The current as a function of $\gamma$ for driving biases $f=0.1,0.2,0.3$ and $0.4$ (from bottom to top curve) with $\Delta=2$, and $N=40$. The dashed lines correspond to $\av{J}_{\Delta=0}$ the non-interacting analytic result~\cite{znidaric10a,znidaric11a}.} \label{fig4}
\end{figure}

For a given $f$ there is an optimal dephasing rate $\gamma_{\rm opt}$ that maximizes the current, as shown in \fir{fig4}. For the parameters of this figure the enhancement at weak driving and the optimal dephasing rate is quite significant, e.g. $\approx37\%$ at $f=0.1$. As the driving increases so does the optimal dephasing rate $\gamma_{\rm opt}$, as well as the enhancement of the current, the latter being of several orders of magnitude for $f\rightarrow1$. For $\gamma > \gamma_{\rm opt}$ the current is reduced because the $\mathcal{L}_d$ contribution to \eqr{master_eq} dominates over the coherent hopping terms and progressively freezes out the dynamics due to the Zeno effect~\cite{breuer02}. In fact for increasingly large $\gamma$ the NESS current converges to the exact $\Delta=0$ solution with dephasing~\cite{znidaric10a,znidaric11a}
\begin{equation}
\av{J}_{\Delta = 0} = -\frac{2f}{\frac{\Gamma}{4}+\frac{4}{\Gamma}+(N-1)\gamma}.
\end{equation}
This convergence, shown in \fir{fig4}, thus indicates that the interaction strength $\Delta$ becomes irrelevant for very large dephasing rates.

\subsection{Scaling with the system size}
To show that the transport enhancement is not restricted to small chains, we analyze the scaling of the optimal dephasing rate $\gamma_{\rm opt}$ with $N$ for weak driving $f=0.1$. The results are presented in \fir{fig5}. Although $\gamma_{\rm opt}$ decreases as $N$ increases, extrapolations with simple trial functions (of which an exponential decay, shown in \fir{fig5}, gives the best description) indicate that even when $N\rightarrow\infty$, $\gamma_{\rm opt}$ remains finite. However, since the density imbalance between the boundaries of the chain $\Delta n$ is bounded, $\langle J\rangle\rightarrow0$ in the thermodynamic limit; see also the inset of \fir{fig5}, which shows that $\langle J\rangle_{\text{opt}}\rightarrow0$ as $N\rightarrow\infty$. Nevertheless, the existence of a finite $\gamma_{\text{opt}}$ in the thermodynamic limit indicates that even in the linear response regime, the transport can be enhanced by environmental coupling in systems of \emph{any} finite size. For stronger driving both the current enhancement and $\gamma_{\rm opt}$ become larger, as shown in \fir{fig4}, and the range of beneficial dephasing rates broadens. So the dephasing-enhanced transport should emerge in mesoscopic systems even for weak driving.

\begin{figure}
\includegraphics{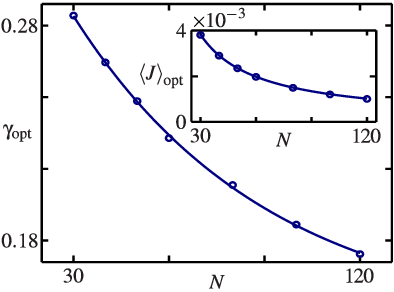}
\caption{Optimal dephasing $\gamma_{\text{opt}}$ for different sizes of the system, at $f=0.1$, $\Delta=2$ and $\Gamma=1$ (circles). The solid line corresponds to the exponential decay $\gamma_{\text{opt}}=\gamma_{\text{opt}}^{\text{TL}}+a\exp{(-N/b)}$, with $a=0.252$, $b=85.9$ and $\gamma_{\text{opt}}^{\text{TL}}=0.109$. Inset: Optimal current $\langle J\rangle_{\text{opt}}$ as a function of $N$. The solid line corresponds to the power law decay $\langle J\rangle_{\text{opt}}=aN^{-b}$, with $a=0.0948$ and $b=0.94$; as $N\rightarrow\infty$, $\langle J\rangle_{\text{opt}}\rightarrow0$.} \label{fig5}
\end{figure}

\section{Enhancement mechanism} \label{mechanism}
We now discuss the physical mechanism underlying NDC and the dephasing enhanced transport. Specifically, we show that these effects arise due to an interplay between the eigen-structure of the strongly-interacting chain Hamiltonian and the boundary driving. As illustrated in \fir{fig6}(a) for a small but representative system size with very strong interactions $|\Delta|\gg1$, the eigen-spectrum consists of nearly flat high-energy bands of bound states of low conductivity, separated by gaps of order $|\Delta|$ from more mobile bands of states with lower energy. As we shall now show boundary driving preferentially populates only the most energetic bound states resulting in an insulating NESS. The introduction of dephasing then induces transitions to mobile current-carrying bands, thereby enhancing the conductivity. 

To see this is more detail it is instructive to first consider maximally biased driving $f=1$ in the extreme $|\Delta| \rightarrow \infty$ limit. In this case configuration states such as $\ket{10110 \cdots 011}$, where the particle occupancy on each site of the chain is explicitly specified, are exact eigenstates of the Hamiltonian. A key property of the boundary driving is that it only incoherently connects configurations within a quadruplet of states $\ket{0 {\bf x} 0},\ket{0 {\bf x} 1} ,\ket{1 {\bf x} 0} ,\ket{1 {\bf x} 1}$ for any value of $f$, where ${\bf x}$ is any length $N-2$ occupancy bit string and thus defines each quadruplet. This is illustrated in \fir{fig6}(b). Consequently, if ${\bf x}$ has $(n-1)$ 1's the driving couples states within the total particle number sectors $n-1, n,$ and $n+1$. This structure constrains the evolution caused by the driving processes to shuffling population between states in these isolated quadruplets. At $f=1$ there is one configuration $\ket{1 {\bf x} 0}$ within each quadruplet which, owing to it having a particle on the leftmost site and a vacancy on the rightmost site, is entirely decoupled from the driving (i.e. there is no action of the driving on such a configuration), while also being the sink for all driving transitions; see \fir{fig6}(b). The effect of this incoherent evolution alone is thus to eventually drive all the population among the quadruplet of states into this {\em dark-configuration}. Of particular relevance are the dark configurations 
\begin{equation} \label{bn}
\ket{B_n} = \kets{\overbrace{111\cdots 111}^n\underbrace{000\cdots 000}_{N-n}},
\end{equation}
which possess an $n$ particle domain pinned to the left boundary. For each particle number sector $n$ the state $\ket{B_n}$ is separated from other configurations by an energy gap $O(|\Delta|)$, akin to a domain binding energy. 

\begin{figure}[t]
\includegraphics{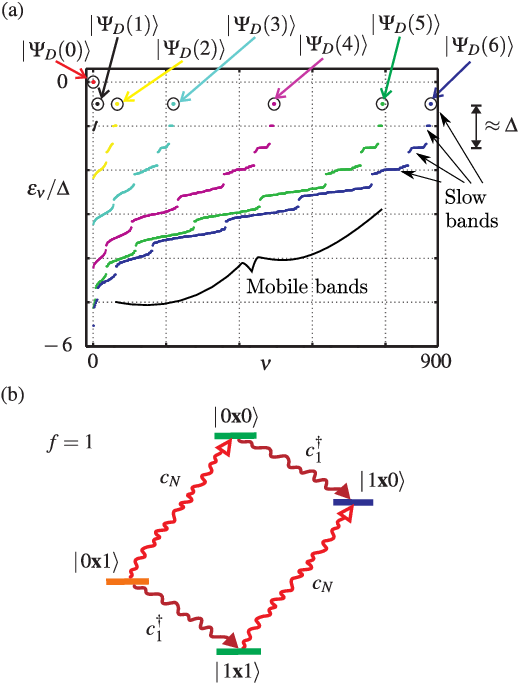}
\caption{(a) The energy eigenspectrum of the spinless fermion Hamiltonian governing the chain for $N=12$ and $\Delta = 10$. Energies $\epsilon_\nu$ have been shifted by $\frac{1}{4}\Delta(N-1)$ so that the state $\ket{\Psi_D(0)} = \ket{00 \dots 00}$ has zero energy. The spectrum includes contributions from all particle number sectors $n = 0, \dots, 12$; since the spectra arising from $n=0, \cdots, 5$ are identical to those of $n=7, \dots, 12$, only the former are shown in addition to $n=6$. The highest lying eigenstate for each sector $\ket{\Psi_D(n)}$ is highlighted and seen to be isolated by a gap of $O(\Delta)$ from eigenstates composed of break-away configurations (states with the outermost particle of the domain breaking away, or a hole propagating through the domain). The braces give an indication of the location of mobile and flattened bands for $n=6$. (b) For any $f$ the driving alone incoherently connects a quadruplet of configurations $\ket{0 {\bf x} 0},\ket{0 {\bf x} 1} ,\ket{1 {\bf x} 0} ,\ket{1 {\bf x} 1}$. The situation for $f=1$ is shown where the driving can be seen to only pump into the so-called dark configuration $\ket{1 {\bf x} 0}$.}
\label{fig6} 
\end{figure}

As we move to the limit of finite, but strong interactions $|\Delta| \gg 1$, hopping between configuration states results in each configuration $\ket{B_n}$ giving rise to an eigenstate $\ket{\Psi_D(n)}$ of bound particles. For each particle number $n$, $\ket{\Psi_D(n)}$ is the highest state in the eigenspectrum, as indicated in \fir{fig6}(a). Its properties are readily determined by treating hopping as a perturbation. Specifically, to lowest-order in $|2\Delta|^{-1}$ hopping hybridizes, across an energy gap of $\Delta$, the state $\ket{B_n}$ with the break-away configuration
\begin{equation} \label{b-a}
\kets{\overbrace{111\cdots 11}^{n-1}01\underbrace{00\cdots 000}_{N-n-1}},
\end{equation}
where the outermost particle of the domain has escaped. As discussed in Appendix \ref{appendix1}, the hybridization of $\ket{B_n}$ with more distant break-away configurations decays exponentially with the distance from the domain wall with a length scale $\xi \sim 1/\ln(|2\Delta|)$. Crucially almost all of these break-away configurations are dark to the driving like $\ket{B_n}$. Only the configurations where either a hole or particle has reached the boundary couple to the driving and their amplitude is exponentially suppressed by this localization. 

The emergence of an insulating NESS in the strongly-interacting regime at $f=1$, having a particle domain of size $N/2$, and thus the domain wall farthest from the boundaries, follows from the combination of two results in the perturbative approach: (i) As $n \rightarrow N/2$, $\ket{\Psi_D(n)}$ becomes an exponentially close approximation to a dark state of $f=1$ driving, with increasing $N$; (ii) The boundary driving at $f=1$ preferentially populates $\ket{\Psi_D(n)}$ leaving a NESS that is well approximated by a statistical mixture
\begin{equation} \label{state_f1}
\rho = \sum_{n=0}^N p_n \ket{\Psi_D(n)}\bra{\Psi_D(n)}, 
\end{equation}
with the probability $p_n$ exponentially peaked at $n=N/2$. In Appendix \ref{appendix1} points (i) and (ii) are shown to arise for sizes $N\geq4$.

The existence of the insulating NESS described by Eq.~\eqref{state_f1} is only possible in the absence of dephasing processes along the chain. Since local dephasing on each site, given in  Eq.~\eqref{dephasing}, does not commute with the chain Hamiltonian $H$ this noise process induces incoherent transitions between many-body energy eigenstates of the system. This is characterized by an energy dissipation rate
\begin{equation} \label{dissipation_rate}
\frac{dE_{\gamma}}{dt}=-2\gamma\sum_j\langle c^\dagger_j c_{j+1} + \textrm{h.c.}\rangle,
\end{equation}
dependent on the kinetic energy of the state, and enables population to escape from the approximate dark states $\ket{\Psi_D(n)}$ to the mobile bands of scattering states. It is this effect which breaks the localization in $f=1$ insulating NESS and significantly enhances the current. An optimal dephasing rate $\gamma_{\text{opt}}$ emerges due to the competition between these dephasing induced transitions and the degradation of mobility of the scattering states by dephasing through the Zeno effect. An increase in $\gamma_{\text{opt}}$ with $|\Delta|$ is observed since a larger energy dissipation is needed to overcome the gap.

When reducing $f$ slightly from the $f=1$ limit a small backward pumping process appears in addition to the dominant forward pumping of particles. Since particles are then injected and ejected by the driving at both boundaries all the states of every quadruplet become populated and there are no longer dark configurations decoupled from the driving. A finite current is therefore established in the NESS as the population in the approximate dark states $\ket{\Psi_D(n)}$ diminishes. Nonetheless the picture of dissipation from majority occupied bound states to higher mobility scattering states still applies. However, in the linear response regime, where $f \rightarrow 0$ and the NESS is a diffusive conductor rather than an insulator, it is not \textit{a priori} obvious that additional dissipation induced by dephasing will be beneficial to transport. Yet as seen in \fir{fig4} and \fir{fig5} an enhancement of the current due to dephasing for chains of any finite size is observed for $f>0$. This behaviour suggests that even in this case, where the high-energy bound states like $\ket{\Psi_D(n)}$ are only marginally populated by the driving, additional incoherent transitions bridging the numerous gaps in the spectrum to the mobile bands still enhance transport. To help further unravel the processes behind dephasing enhancement and NDC we describe a simple toy model in the following section. This model not only reproduces the basic features of these effects in a concrete analytically tractable way, but also reveals that the physical mechanism underlying both effects is the same.  

\section{Toy model} \label{toy}

We have seen that numerous approximate dark states, whose occupation is favoured by $f=1$ driving, cause the stationary state to become insulating. Remaining in the strongly interacting limit we now wish to isolate the effect of such bound states on the transport for all drivings $f$, i.e. on the complete current-driving profile. To do so we construct a simple toy model, whose structure is motivated by considering the half-filled domain state $\ket{B_{N/2}}$ and its corresponding break-away configurations which eventually connect it to the boundary driving. As such the toy model is composed of $K$ configuration states $\ket{1},\ket{2},\dots,\ket{K}$ for some size $K>2$. To mimic the interaction binding energy of $\ket{B_{N/2}}$, we distinguish the configuration $\ket{K}$ by elevating it in energy by $\Delta$ above the set of otherwise degenerate configurations $\ket{1},\ket{2},\dots,\ket{K-1}$ which model break-away states like $\ket{11\cdots10100\cdots0}$, $\ket{11\cdots10010\cdots0}$ etc. In addition, the states $\ket{1},\ket{2},\dots,\ket{K}$ are also coherently coupled to their neighbours via ``hopping" processes given by 

\begin{figure}
\includegraphics{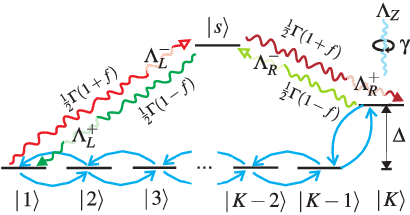}
\caption{The schematic of the toy model and its various processes and properties. These include the nearest-neighbour coherent hopping between the set of states $\ket{1}, \ket{2}, \dots, \ket{K}$, an energy offset $\Delta$ for state $\ket{K}$, incoherent transitions $\ket{1} \leftrightarrow \ket{s}$ and $\ket{K} \leftrightarrow \ket{s}$ between the boundary states and the auxiliary state $\ket{s}$, and dephasing at a rate $\gamma$. The rates for the incoherent driving transitions $\half\Gamma (1 \pm f)$ are also listed.}
\label{fig7} 
\end{figure}

\begin{equation}
 H_t = \half \sum_{k=1}^{K-1}(\ket{k}\bra{k+1} + \textrm{h.c.}). \nonumber
\end{equation}
The current operator for the model then follows as
\begin{equation}
J = -i\sum_{k=1}^{K-1}(\ket{k}\bra{k+1} - \textrm{h.c.}), \nonumber
\end{equation}
which measures the flow within the coherently connected configurations $\ket{1},\ket{2},\dots,\ket{K}$. To model the driving in the full system, which incoherently connects one particle number sector to another, we introduce an auxiliary state $\ket{s}$ whose function is simply to be an intermediary. The jump operators describing the driving then take the form 
\begin{equation}
L_{L}^\pm=\sqrt{\Gamma(1\mp f)/2}\,\Lambda_{L}^\pm,\quad L^\pm_{R}=\sqrt{\Gamma(1\pm f)/2}\,\Lambda_{R}^\pm, \nonumber
\end{equation}
where $\Lambda_L^- = \ket{s}\bra{1}$ and $\Lambda_R^+ = \ket{K}\bra{s}$, with $\Lambda_L^+ = (\Lambda_L^-)^\dagger$ and $\Lambda_R^- = (\Lambda_R^+)^\dagger$. Thus, via $\ket{s}$, the driving incoherently induces transitions between the boundary configurations $\ket{1}$ and $\ket{K}$ with a bias $f$. At $f=0$ driving in both directions is equal and it is easily confirmed that the NESS is $\rho = \mathbbm{1}/(K+1)$, yielding $\av{J} = 0$ as in the case of the full spinless fermion chain. At the opposite limit, $f=1$, population is asymmetrically driven from $\ket{1} \rightarrow \ket{K}$. To complete the analogy with the full system the toy model also includes dephasing, at a rate $\gamma$, via the jump operator $L_Z = \sqrt{\gamma}\Lambda_Z$ where
\begin{equation}
\Lambda_Z = \mathbbm{1} - 2\ket{K}\bra{K}, \nonumber
\end{equation}
whose action is to scramble the phase of any superpositions between $\ket{K}$ and the other configurations. A schematic of the toy model illustrating all the coherent and incoherent contributions is shown in \fir{fig7}.

\begin{figure}
\includegraphics{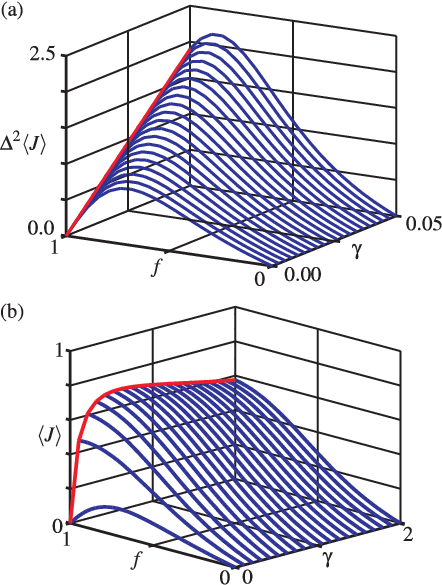}
\caption{(a) The NESS current $\av{J}$, rescaled by $\Delta^2$, for $K=20$, as a function of the driving $f$ and dephasing $\gamma$, described by the approximation in \eqr{eq:toy_model}. The behaviour at $f=1$ with $\gamma$ is emphasized by the additional (red) line. (b) The exact current-driving profiles of the toy-model as a function of $\gamma$ with $\Delta=2$ and $K=20$.}
\label{fig8} 
\end{figure}

Since it is inspired in the $|\Delta|\gg1$ regime, the toy model does not embody the entire physics of the full system. Nevertheless it does capture the essential features of NDC and dephasing enhancement when $|\Delta|\gg1$. Fundamentally, the same mechanisms observed in such a limit apply to the more complex full system when considering its entire eigen-spectrum, even when $|\Delta|\rightarrow1$. 

The current $\av{J} = \textrm{tr}(J\rho)$ of the NESS $\rho$ can be solved analytically for the toy model as a function of $f,\gamma$ and $\Delta$, although the complete expression is lengthy. Since the model was motivated by the perturbative limit $|\Delta| \gg 1$ the physically relevant part of this result is found by keeping only the lowest order terms in $\Delta^{-1}$. This gives
\begin{eqnarray}
\av{J} &\approx& \frac{(K-1)\Big(8\gamma f + (1-f)f\Gamma\Big)}{(K+1)- 2(K-2) f+ (K-1)f^2}\left(\frac{1}{\Delta}\right)^2. \label{eq:toy_model}
\end{eqnarray}
In \fir{fig8}(a) the current-driving profile $\av{J}$, rescaled by $\Delta^2$, is plotted for $K=20$. Two key features emerge from this result. First, for $\gamma = 0$ the $(1-f)f$ in the numerator of \eqr{eq:toy_model}, which enforces zero current at the $f=0$ and $f=1$, causes $\av{J}$ to display negative differential conductivity (NDC). Moreover, the expansion involves only even powers of $\Delta^{-1}$, showing that this behaviour is independent on the sign of $\Delta$. Second, dephasing enhancement of $\av{J}$ is evident from the linear $\gamma$ term in \eqr{eq:toy_model} which eventually destroys the NDC effect. However, since $\gamma$ and $\Gamma$ appear only linearly, this lowest order expression is only valid for weak dephasing and coupling to the boundary reservoirs. In particular the dephasing degrading behaviour expected for large $\gamma$, due to the Zeno effect, is not described by \eqr{eq:toy_model}. The expression is also only valid for $K \geq 3$ because NDC is not seen for $K=2$; having no direct coherent coupling between the boundary configurations where driving occurs, i.e. $\ket{1}$ and $\ket{3}$ for $K=3$, is essential for NDC to emerge. This is similar to how NDC is only seen for $N \geq 4$ in the full system, like in \fir{appfig1}(b) of Appendix \ref{appendix1}. In \fir{fig8}(b) the exact current-driving profile for the toy model is shown for $\Delta = 2$ as a function of moderate $\gamma$'s, beyond the applicability of \eqr{eq:toy_model}. This confirms the wider similarity of the response of the toy model to that observed in the full spinless-fermion system. 

\begin{figure}
\includegraphics{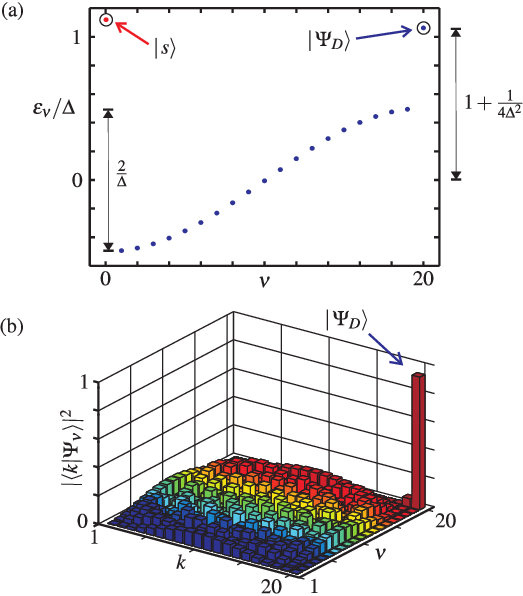}
\caption{(a) The spectrum $\epsilon_\nu$ of the Hamiltonian for the toy model for $K=20$ and $\Delta = 2$. Eigenstate $\nu = 0$ is the highlighted auxiliary state $\ket{s}$
whose energy has been set arbitrarily. The highest-lying eigenstate $\nu = K$ is also highlighted and is the approximate dark-state of the model $\ket{\Psi_D}$. Beneath this state, separated by a gap $O(|\Delta|)$, is a band of eigenstates $\ket{\Psi_\nu}$ split by the small hopping. (b) The probability distributions $|\braket{k}{\Psi_\nu}|^2$ of the eigenstates over the configurations $\ket{k}$ are shown. The band of eigenstates are seen to be delocalized over the configurations $\ket{1}, \ket{2}, \dots, \ket{K-1}$ and expunged from the boundary configuration $\ket{K}$, while the characteristic of a bound state $\ket{\Psi_D}$ is predominately peaked at $\ket{K}$ with exponential tail into the bulk.}
\label{fig9} 
\end{figure}

Given their similar response, the toy model provides a tractable means of unravelling the origins of NDC and dephasing enhancement in the full many-body system. The eigenspectrum of the toy model is plotted in \fir{fig9}(a). By construction we see that it mimics some of the features shown in \fir{fig9} for an individual number sector of the spinless fermion system. Specifically, there is a high-lying eigenstate $\ket{\Psi_D}$, separated by a gap $O(|\Delta|)$ from a dense band of eigenstates. In \fir{fig9}(b) the eigenstates of this band are seen to be delocalized over the bulk of the system excluding the boundary configuration $\ket{K}$. In contrast, for a given size $K$, the eigenstate $\ket{\Psi_D}$ has the form 
\begin{equation}
\ket{\Psi_D} \approx \sum_{k=0}^{K-1} |2\Delta|^{-k} \ket{K-k}, 
\end{equation}
to within $O(|2\Delta|^{-2K})$, as seen in \fir{fig9}(b). As $K \rightarrow \infty$ such an exponentially decaying wave function is simply the discrete analogue of the well known bound state of a 1D $\delta$-potential~\cite{griffiths05}. Given that the amplitude for the left boundary configuration $\ket{1}$ scales as $|2\Delta|^{1-K}$, we see that $\ket{\Psi_D}$ becomes exponentially close, with increasing $K$, to being a dark state of the driving when $f=1$. 

The driving at $f=1$ exclusively pumps into the configuration $\ket{K}$ whose dominant overlap is with $\ket{\Psi_D}$. Consequently so long as $\gamma = 0$ population gets progressively trapped in this dark state giving rise to $\av{J}\approx 0$, characteristic of an insulating NESS. Remaining at $f=1$ and switching on a non-zero dephasing directly decoheres the exponentially decaying superposition within $\ket{\Psi_D}$. This is equivalent to the coherent trapping of population, caused by the energetic gap, being bypassed by dephasing induced incoherent transitions connecting $\ket{\Psi_D}$ directly to the delocalized band of eigenstates. Current flow in the system is thus made possible via the ensuing non-stationary mixture of these eigenstates. Further increases in dephasing eventually degrades the current once the monotonically decreasing mobility of the delocalized eigenstates, caused by the Zeno effect, outweighs the flux of population escaping from $\ket{\Psi_D}$. Since the toy model only has a single approximate dark state $\ket{\Psi_D}$ it confirms that its existence, at one isolated point $f=1$ and $\gamma=0$, is alone enough to make the current-driving profile for $0 \leq f \leq 1$ exhibit NDC and $\gamma \geq 0$ exhibit dephasing enhancement.

Another key insight from the toy model is that the emergence of a non-zero current from the insulating point $f=1$ and $\gamma=0$, involves identical physics either when $\gamma$ is increased slightly from zero, or when $f$ is reduced slightly from unity. Examining \eqr{eq:toy_model} at $f=1$ shows that $\av{J}\Delta^2 = 2(K-1)\gamma$, while for $\gamma=0$ an expansion about $f=1$ the current is $\av{J}\Delta^2 = \frac{1}{4}(K-1)(1-f)\Gamma$ to lowest order. This suggests a correspondence
\begin{equation}
\gamma = \half(1-f)\frac{\Gamma}{4}. \label{eq:gamma_f}
\end{equation}
Consequently, a slight decrease of driving or increase of dephasing induce the same decoherence process that enhances the transport of the otherwise insulating state. A further indication of this equivalence, focused on the decay in time of the coherences between states $\ket{K-1}$ and $\ket{K}$, is presented in Appendix \ref{appendix2}.

While successful in describing the effective single-particle aspects of the full spinless fermion system the toy model fails to describe several important features that are hallmark of genuine many-body physics. First, in the strongly driven $f=1$ limit at large dephasing the toy model does not display diffusive transport, like that observed for the full system in \fir{fig3}. This is expected since aside from configuration $\ket{K}$ the toy model is an ordered homogeneous tight-binding system. Second, when there is no dephasing, the full system is known to exhibit a non-equilibrium phase transition from diffusive to ballistic transport at weak driving~\cite{benenti09b} as $|\Delta| \rightarrow 1$. As the toy model was constructed to mimic the $|\Delta| \gg 1$ limit it does not capture this many-body property of the full system. In the next section we investigate the interplay of dephasing on this transition in the full system.  

\section{Signature of a non-equilibrium phase transition} \label{phase}

The results discussed in Section \ref{enhancement} demonstrate the existence of two transport regimes in the system with different responses to moderate dephasing: degradation for weak interactions and enhancement for strong interactions. We now discuss the transition between the two transport regimes, characterize the critical interaction strength, analyze the correlations through the system for each regime, and show that the same regimes of response remain even when the integrability of the system is broken.

\begin{figure}
\includegraphics{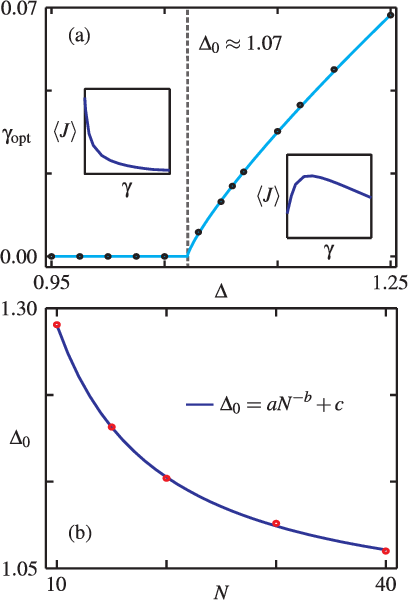}
\caption{(a) The optimal dephasing rate $\gamma_{\text{opt}}$ as a function of $\Delta$, for $N=40$, $f=0.1$ and $\Gamma=1$. The circles indicate TEBD results, and the solid line corresponds to the fitted function $\gamma_{\text{opt}} \propto (\Delta-\Delta_0)^{\beta}$, with $\beta=0.819$ and $\Delta_0=1.07$. The inset plots show the generic behaviour of $\av{J}$ with $\gamma$ above and below $\Delta_0$. (b) The scaling of $\Delta_0$ with $N$ for $f-0.1$ and $\Gamma=1$, along with the fitted power-law shown, where $a=3.906$, $b=1.066$ and $c=0.995$.} \label{fig10}
\end{figure}

\subsection{Transition at $|\Delta|=1$}

The interaction strength $|\Delta|=1$ is of particular significance to the Hamiltonian of Eq. \eqref{hami}. At zero-temperature and in the absence of magnetic field, it separates the gapless and magnetically ordered (ferro- and antiferromagnetic) gapped equilibrium phases. More generally, it divides a continuous eigenspectrum for $|\Delta| < 1$ from one with numerous gaps for $|\Delta|>1$~\cite{sutherland05,takahashi99}, as illustrated in \fir{fig6}(a). In the system driven by the jump operators of Eq. \eqref{jump_driving}, it was previously observed that $|\Delta|\approx 1$ also separates ballistic and diffusive transport regimes for weak driving~\cite{benenti09b}. However, it is not \textit{a priori} clear that $|\Delta|=1$ necessarily separates the regimes of transport degradation and enhancement by dephasing. We now show that this is indeed the case. 

In \fir{fig10}(a) the optimal dephasing rate $\gamma_{\rm opt}$ is shown for weak driving as a function of $\Delta$. A threshold of $\Delta_0\approx 1.07$ is apparent where for $|\Delta| < \Delta_0$, $\gamma_{\text{opt}}=0$, indicating dephasing-degraded transport, while for $|\Delta|>\Delta_0$, $\gamma_{\text{opt}}$ is non-zero, indicating dephasing-enhanced transport, and increases monotonically with $|\Delta|$. The latter behavior is a consequence of the enhancement mechanism. As the interaction strength increases, so do the gaps between flattened and mobile bands, so a larger energy dissipation is required to populate the latter and increase the current. The value of $\Delta_0$ is size-dependent, and a scaling analysis shown in \fir{fig10}(b) demonstrates that, to good approximation, $\Delta_0=1$ separates the two transport regimes in the thermodynamic limit. 

From the results discussed above it is tempting to associate the nature of the ground state, namely gapped or gapless, with a certain type of particle transport or response to dephasing. This impression is misleading, as can be understood by considering a homogeneous on-site potential $B\sum_jn_j$. This potential shifts the equilibrium quantum critical points~\cite{takahashi99}, but leaves the particle transport of the steady state unaltered. The latter occurs because the Hamiltonian and the current operator are particle-conserving, so the sectors of different particle number are only incoherently connected due to the driving and the NESS is block diagonal. The internal structure of the various particle number sectors, unmodified by the homogeneous on-site potential, is thus what determines the steady-state transport through the chain, not the relative positions of the different sectors within the eigen-spectrum~\cite{mendoza13}. This shows that a qualitative change of the ground state does not imply a change of non-equilibrium properties. Instead, it is the overall structure of the eigen-spectrum which leads to the NDC and dephasing-enhanced transport effects at strong interactions, as discussed in Sections \ref{mechanism} and \ref{toy}.   

\subsection{Correlations and dephasing}
Similar to equilibrium phases, the transition between the two different transport regimes, namely of current degradation and enhancement by dephasing, can also be distinguished from the correlations through the system. First we consider a typical two-point density-density correlation function
\begin{equation} \label{correlations}
C_{ij}=\langle n_i n_j\rangle-\langle n_i \rangle\langle n_j \rangle,
\end{equation}
with sites $i$ and $j$ symmetrically positioned around the centre of the chain. This is conveniently represented as $C(r)$ where $r=|i-j|/N$ is the fractional separation of the points for the system size $N$. In \fir{fig11}(a) $C(r)$ is plotted for different sizes $N$ in the strongly interacting regime for a moderate dephasing rate. Finite correlations are seen to exist even for a large fractional separation $r$. Although these correlations decay with $r$, the fraction of the system they span increases with $N$. This property has been argued, for $\gamma = 0$ and $|\Delta|>0.91$, to be evidence that the NESS possesses genuine long-range order even in the thermodynamic limit~\cite{prosen10}. Here our results indicate that this long-range order persists even in the presence of moderate dephasing, and so correlations similar to those at $\gamma = 0$ exist in the dephasing enhanced NESS.  The situation for strong dephasing, shown in \fir{fig11}(b), is markedly different. Now correlations are smaller and diminish faster with $r$ than at weaker dephasing rates. In addition, the fraction of the system over which the correlations extend diminishes as $N$ increases, a behavior previously observed for large dephasing rates $\gamma$ independently of the interaction strength $|\Delta|$~\cite{znidaric10b,znidaric10a}. So like the current and density profiles, the two-point correlation functions in the strongly-interacting regime become increasingly similar to those of the weakly interacting regime for large dephasing, washing out the transition.

\begin{figure}
\includegraphics{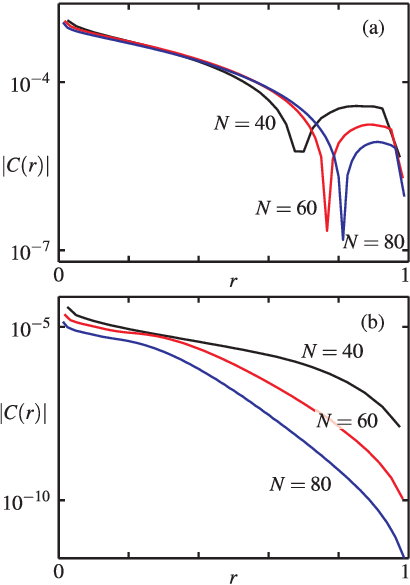}
\caption{The NESS density-density correlations $|C(r)|$ across the chain as a function of the fraction distance $r$ about the centre for $\Delta = 2$ and $f=0.1$. In (a) a moderate dephasing $\gamma = 0.05$ was included. The correlations are long-ranged and extend over a larger fraction when increasing $N$. This evidenced by the kink, coming from $C(r)$ changing sign, moving to larger $r$. In (b) a strong dephasing $\gamma = 1$ was present and the correlations are short-ranged as shown by their diminishing size and extent with increasing $N$.}
\label{fig11} 
\end{figure}

The signature of the non-equilibrium transition between weakly- and strongly-interacting regimes with moderate dephasing, already suggested by two-point correlation functions, can be refined by adopting a more general measure of correlations. Specifically we compute the entropy~\cite{zwolak04,prosen09,znidaric08}
\begin{equation}
S=-\sum_\alpha\lambda^2_\alpha\log_2\lambda^2_\alpha 
\end{equation}
of the Schmidt coefficients $\lambda_\alpha$ arising when the full NESS density operator $\rho$ is factorized into two half-chains as 
\begin{equation}
\rho = \sum_\alpha \lambda_\alpha O^A_\alpha O^B_\alpha,
\end{equation}
where $O^A_{\alpha}$ and $O^B_{\alpha}$ are Hilbert-Schmidt orthogonal operators for the two halves. Both quantum and classical correlations between the two halves of the chain are quantified by $S$ and it is readily accessible from the TEBD numerics. For zero dephasing and weak driving \fir{fig12} shows that $S$ peaks at $\Delta \approx 1$. This indicates that a significant elevation in correlations occurs as the NESS reorganises itself across the expected non-equilibrium phase transition between ballistic and diffusive transport in this region. 

\begin{figure}
\includegraphics{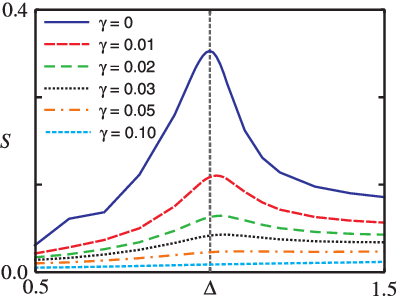}
\caption{The half-chain entropy $S$ as a function of $\Delta$ for different $\gamma$'s, $f=0.1$ and $N=40$. The Schmidt coefficients $\lambda_{\alpha}$ were normalized so the largest coefficient $\lambda_1=1$.} \label{fig12}
\end{figure}

In \fir{fig12} we also show that for finite dephasing the entropy $S$ monotonically decreases with $\gamma$, but interestingly the peak is maintained for moderate rates, shifting slightly to larger $\Delta$. Therefore the abrupt on-set of dephasing enhancement shown in \fir{fig10} occurs when the many-body correlations are strongest, indicating that it is an effect which applies far beyond the effective single-particle picture of the toy model. This distinguishing feature in $S$ is progressively washed out by increasing dephasing as the system becomes diffusive for all $\Delta$~\cite{znidaric10a,znidaric11a,znidaric10b}. Despite this, the response of the system to the dephasing processes is seen to offer an alternative and clear signature of the underlying non-equilibrium phase transition between two qualitatively different steady states. 
  
\subsection{Enhancement and integrability}  
The Hamiltonian in Eq.~\eqref{hami} describing the full system is integrable~\cite{sutherland05}, so it could be considered that the dephasing enhancement observed in the present work is an artifact of this property. To show this is not the case, we obtain the NESS of the system when adding a staggered local potential 
\begin{equation}
H_{\text{s}}=B\sum_{j=1}^N(-1)^j n_j 
\end{equation}
which breaks its integrability. For $\gamma = 0$ this has the effect of turning the system into a diffusive conductor in the gapless regime $|\Delta|<1$ for any driving, while not affecting the existence of NDC at large drivings for $|\Delta|>1$~\cite{benenti09b}. In \fir{fig13}(a) we see that for weak interactions dephasing monotonically decreases the current, while for strong interactions we confirm, for a variety of field strengths $B$, that dephasing-enhanced transport occurs, as shown in \fir{fig13}(b). Thus the existence of NDC and dephasing enhancement, characterizing the non-equilibrium phase transition between weakly- and strongly-interacting regimes, is independent of integrability. Both effects arise as long as the eigen-structure of the system possesses the features discussed in Sections \ref{mechanism} and \ref{toy}, which for the case considered in this work is valid even if the integrability is broken, as shown in the inset of \fir{fig13}(b).

\begin{figure}
\includegraphics{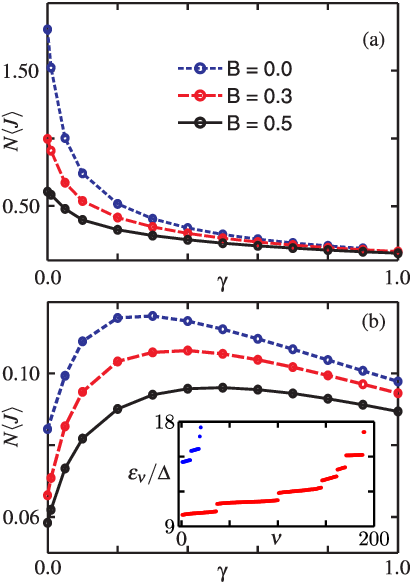}
\caption{The current through the system for several staggered potential strengths $B$, for $N=40$, $f=0.1$ and $\Gamma=1$, with (a) $\Delta=0.5$ and (b) $\Delta=2.0$. Inset. Energy eigen-spectrum for one (blue) and two (red) excitations of a chain of $N=20$, with staggered potential $B=2$ and interaction $\Delta=10$.} \label{fig13}
\end{figure}   

From the eigen-structure of the non-integrable system, the most notable features of the dephasing-enhanced transport can be understood. Namely, the increase of the optimal dephasing rate $\gamma_{\text{opt}}$ with $B$ results from the splitting of the energy bands due to the staggered potential, as shown in \fir{fig13}(b) (compare to \fir{fig6}(a)). This split emerges from different energy shifts of the eigen-states of the system, which depend on their spatial distribution\footnote{The simplest case illustrating this point is that of one excitation in an even chain with $\Delta\rightarrow\infty$; if $B=0$ the states $\ket{10\cdots0}$ and $\ket{0\cdots01}$ are degenerate, while if $B>0$ they experience negative and positive energy shifts, respectively.}. Due to the band splitting and the emergence of new energy gaps, the states of wide bands of lowest energy become harder to populate, so a larger energy dissipation is required to induce transitions towards them, and thus to enhance the transport. In addition, note that the staggered potential also flattens all the energy bands, reducing the conductivity through the chain.

\section{Conclusions} \label{conclu}
We have presented a detailed study of the effects of dephasing on the transport properties of a boundary driven one-dimensional interacting spinless fermion chain. The appearance of a cooperative many-body NESS exhibiting NDC at strong interactions provides a previously unexplored form of dephasing enhanced transport, distinct in origin from earlier examples in non-interacting systems. Using a toy model for the very-strongly interacting regime, we isolated the minimum requirements for observing NDC and dephasing enhanced transport. These consist of the emergence of a gapped eigen-structure with bands of eigenstates possessing different mobility, and the transitions between states of different bands induced by incoherent processes. At maximal driving and no dephasing, approximate dark states are preferentially populated, inducing an insulating steady state. The introduction of dephasing populates mobile bands due to energy dissipation, turning the system into a diffusive conductor. A similar mechanism occurs even in the linear response regime of very large systems, leading to a significant enhancement of the current.

While our discussion of the transport properties has been focused on the NESS of the driven interacting system, it is also highly relevant for transient dynamics. Specifically, since the growth of a particle domain requires the propagation of holes toward the left boundary, the increasing suppression of this process in each successive particle number sector $n$ leads to an exponentially slow convergence to the NESS~\cite{benenti09b}. The transient current is effectively suppressed even for a small domain $n>5$ and this is the physical reason why our numerical calculation of the NESS for $f=1$ is limited to relatively small systems when $\gamma =0$. As such the existence of approximate dark states $\ket{\Psi_D(n)}$ at $f=1$ heavily influences the dynamics far from stationarity and indicates that dephasing enhancement will be significant in the transient regime as well.

For moderate dephasing the different nature of the NESS at weak and strong interactions was revealed by the emergence of large correlations, and reflects an underlying non-equilibrium phase transition in the system. As such dephasing enhancement as well as the NDC for $|\Delta| \rightarrow 1$ are truly many-body phenomena. A recent study has also observed both NDC and dephasing enhancement of heat transport in the same system with strong interactions~\cite{mendoza13}. These effects are also unrelated to integrability, suggesting that our findings will also apply to more realistic strongly correlated systems such as the $t-J$ or Hubbard models.  

Dephasing enhancement is also expected to be found in non-equilibrium systems with a different transport process. The most important example corresponds to that of the expansion on an initially-trapped wave packet~\cite{gobert05,langer09,jesenko11}. Tantalising signs of noise induced breakup of bound states and subsequent increase of the expansion of a strongly-interacting packet have already been observed in a recent cold-atom experiment~\cite{errico12}. The general enhancement mechanism predicted in the present work also provides an explanation for this result when applied to this type of transient dynamics. Furthermore, given the recent advances in experiments on transport of ultracold atomic gases, the prospects of verifying more directly both NDC and dephasing enhancement are promising~\cite{brantut12,stadler12}.

\begin{acknowledgments}
J. J. M.-A. acknowledges Departamento Administrativo de Ciencia, Tecnolog\'{i}a e Innovaci\'{o}n Colciencias for economic support. D.J. and S.R.C. thank the National Research Foundation and the Ministry of Education of Singapore for support. 
\end{acknowledgments}

\appendix

\section{Existence of an insulating state at $f=1$ when $|\Delta|\gg1$} \label{appendix1}

\begin{figure*}[t]
\includegraphics{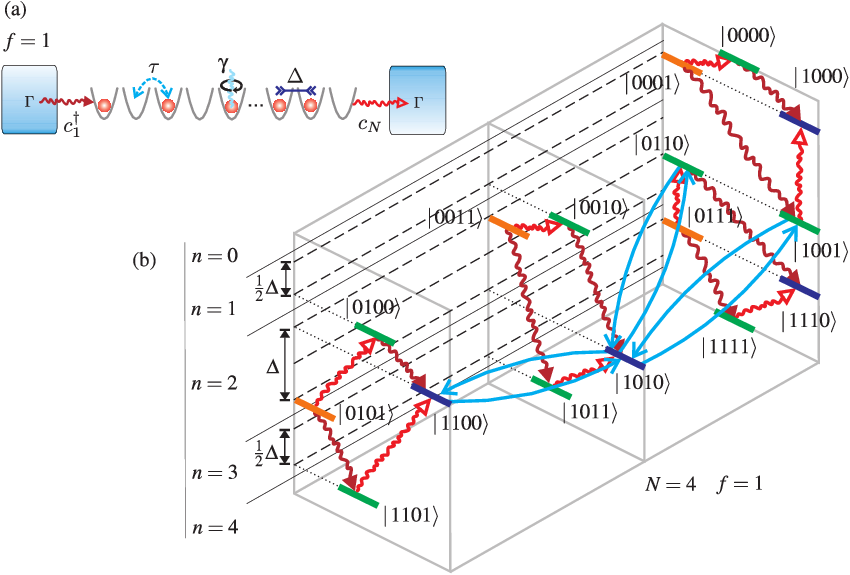}
\caption{(a) A schematic of the system, showing the coherent hopping $\tau$, the nearest-neighbour density-density interaction $\Delta$, dephasing $\gamma$, and the incoherent boundary driving in the strong driving limit $f=1$ where the bias is maximal. In this case the driving incoherently pumps particles into the system at site 1 and ejects them at site $N$, both at a rate $\Gamma$. (b) For $N=4$ sites and $f=1$ the complete set of driving quadruplets and the transitions induced by coherent hopping between them from the half-filled domain state $\ket{1100}$ is shown. The vertical axis displays the particle number sectors and energy gaps between configurations. We observe that for $N=4$ one hop connects $\ket{1100}$ to $\ket{1010}$, which is a dark configuration of its quadruplet, and a further hop is needed before it is coherently connected to a configuration that is not dark to the $f=1$ driving. This disconnection between the half-filled domain configuration and the driving only occurs for $N \geq 4$, and so this is the smallest size which displays NDC.}
\label{appfig1} 
\end{figure*}

Formally a dark state $\ket{\psi}$ of an open quantum system is a pure state which is simultaneously an eigenstate of the Hamiltonian $H\ket{\psi} = E\ket{\psi}$ and a zero-eigenvalue eigenstate of all the jump operators comprising the dissipator $\mathcal{L}(\ket{\psi}\bra{\psi}) = 0$ that describes the noise acting on the system. In this Section we show that deep in the strongly-interacting limit $|\Delta| \gg 1$, with maximal driving $f=1$, as indicated in \fir{appfig1}(a), there exists a state $\ket{\Psi_D}$ which, as the system size $N$ grows, becomes exponentially close to satisfying these requirements and is thus an approximate dark state. As described in Section \ref{mechanism}, the demonstration consists of two parts.

\subsection{Existence of a dark state for maximal driving $f=1$}

As shown in \fir{fig6}(b), the driving scheme of the strongly-interacting limit $|\Delta|\rightarrow\infty$ couples the four eigenstates $\ket{0 {\bf x} 0},\ket{0 {\bf x} 1} ,\ket{1 {\bf x} 0} ,\ket{1 {\bf x} 1}$, establishing a quadruplet defined by the string ${\bf x}$. At $f=1$ the states $\ket{1 {\bf x} 0}$ become dark, of which the $n-$particle domain configurations $\ket{B_n}$ play a prominent role. Namely, for finite but large interaction $|\Delta|$, they weakly hybridize with break-away configurations, resulting in the high-energy bound eigenstates $\ket{\Psi_D(n)}$. In \fir{appfig1}(b) the pattern of incoherent driving transitions and coherent hopping for $N=4$ is illustrated. Note that in this case $\ket{B_2}=\ket{1100}$ has no direct coherent transition to any configurations which couples to the $f=1$ driving. It only couples to the state $\ket{1010}$, which is also a dark configuration. More generally, so long as $N \geq 4$ and $n \geq 2$, the break-away configuration for any domain state $\ket{B_n}$ is also a dark configuration of its own quadruplet.  

We now compute the high-order corrections to the states $\ket{B_n}$ due to hopping to build a perturbative picture of the bound eigenstates $\ket{\Psi_D(n)}$. Since $|\Delta| \gg 1$ it is instructive to do this approximately by focusing on states describing a single break-away particle or hole propagating away from the domain wall at site $n$. These have the form $c_j c^\dagger_k\ket{B_n}$, where $1\leq j\leq n$ and $n< k\leq N$. Since the repeated action of hopping on $\ket{B_n}$ originates around the domain wall, and is also detuned by the gap $\Delta$, we find that hopping mixes in configurations where the particle and/or hole have hopped $x$ times in total, with an amplitude scaling as $O(|2\Delta|^{-x})$. This indicates that the particle/hole propagation through the empty/unit-filled regions is suppressed with its distance from the domain wall. For each $n$, the eigenstate $\ket{\Psi_D(n)}$ is therefore predicted by this particle-hole (ph) picture to have a domain wall that remains exponentially localized at site $n$, within a length scale $\xi \sim 1/\ln(|2\Delta|)$~\cite{benenti09b}, and a deviation $\delta_n(j)$ from the perfect domain configuration $\ket{B_n}$ given by
\begin{equation} \label{eq:deviation}
\delta^{\textrm{ph}}_n(j) =
\left\{
\begin{array}{lr}
 \left(\frac{1}{|2\Delta|}\right)^{2(n-j+1)}, & 1 \leq j \leq n    \\
 \left(\frac{1}{|2\Delta|}\right)^{2(j -n)}, &  n < j \leq N 
\end{array}
\right. .
\end{equation}
The validity of this picture is established by comparing this to the actual density deviation $\delta_n(j)$ of the exact eigenstate $\ket{\Psi_D(n)}$. In \fir{appfig2} this is done for $\ket{\Psi_D(N/2)}$ with $N=12$ and $\Delta=10$, and the agreement between $\delta_n(j)$ and $\delta^{\textrm{ph}}_n(j)$ is seen to be excellent everywhere but the boundaries. 

Since hopping predominately connects $\ket{B_n}$ with dark-configurations of the form $\ket{1 {\bf x} 0}$, the leading order contribution to the amplitude in $\ket{\Psi_D(n)}$ for a configuration which is not dark is $O(|\Delta|^{-\textrm{min}(n,|n-N/2|)})$. This corresponds to whether the hole or particle has the shortest path to the left or right boundary, respectively. In the $|\Delta| \gg 1$ limit we therefore conclude that the eigenstates $\ket{\Psi_D(n)}$, with $0 \leq n \leq N/2$, form a hierarchy of states with $n$, characterised by a decreasing amplitude for any configuration to be coupled to the $f=1$ boundary driving. Eigenstates $\ket{\Psi_D(n)}$ with a domain size $n$ scaling with $N$ thus become exponentially close to being zero eigenstates of the $f=1$ driving with increasing system size. Of these states the one with $n=N/2$, where the domain spans exactly half the chain, has the most suppressed coupling $O(|\Delta|^{-N/2})$ to the driving and is thus the closest approximation of all of them to an exact dark state. Now we show that this is precisely the state preferentially populated by the driving process.

\begin{figure}
\includegraphics{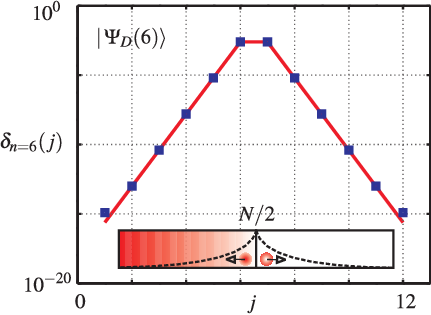}
\caption{For the eigenstate $\ket{\Psi_D(6)}$ for $N=12$ and $\Delta = 10$ the exact density deviation $\delta_n(j)$ from the corresponding boundary domain configuration $\ket{B_n}$ is shown $(\blacksquare)$, along with $\delta^{\textrm{ph}}_n(j)$ predicted by considering single particle/hole propagation (solid line). The inset shows a schematic of the expected exponential localization of the domain wall.}
\label{appfig2} 
\end{figure}

\subsection{Structure of the NESS in the $|\Delta| \gg 1$ limit}

The open dynamics of particle number conserving systems like that considered here have been studied extensively~\cite{burgarth07,benenti09b} with regard to their ergodic and mixing properties, and it has been established that a unique NESS exists for any $f$. The key property ensuring this is that for a finite hopping amplitude every configuration within each particle number sector can be reached from any other, while the incoherent ejection/injection of particles by the driving connects neighbouring sectors. As a result the complete state space of the system can be accessed. Furthermore, the NESS for this open system will be block diagonal in the number sectors. At $f=1$ the approximate dark states $\ket{\Psi_D(n)}$ for each sector $n$ are expected to play a prominent role due to their ability to trap population. This can be better understood by approximating the NESS as a statistical mixture, with probabilities $p_n$, of these eigenstates in each sector as
\begin{equation}
\rho = \sum_{n=0}^N p_n \ket{\Psi_D(n)}\bra{\Psi_D(n)}.
\end{equation}

The purity of the NESS in this approximation, $\textrm{tr}(\rho^2) = \sum_{n=0}^N p_n^2$, is reduced only through mixing between sectors. The probabilities $p_n$ are then determined by demanding that at stationarity there is a detailed balance condition between the incoherent transition rates connecting neighbouring number sectors (see inset of \fir{appfig3}(b)). For sector $n$, assumed to be frozen in the state $\ket{\Psi_D(n)}$, the output transition rates scale with the probability of a hole being at the left boundary as $\Gamma \bra{\Psi_D(n)}c_1c_1^\dagger\ket{\Psi_D(n)} \sim \Gamma |2\Delta|^{-2n}$, and the probability of a particle being at the right boundary as $\Gamma \bra{\Psi_D(n)}c_N^\dagger c_N\ket{\Psi_D(n)} \sim \Gamma |2\Delta|^{-2(N-n)}$. Considering sectors $n-1$, $n$ and $n+1$ we then have the equality of incoming and outgoing transitions in $n$ as
\begin{eqnarray}
p_n\left[\left(\frac{1}{|2\Delta|}\right)^{2n} + \left(\frac{1}{|2\Delta|}\right)^{2(N-n)}\right] = \quad\quad\quad\quad\quad\quad\quad\quad \nonumber \\
\quad\quad p_{n-1}\left(\frac{1}{|2\Delta|}\right)^{2(n-1)} + p_{n+1}\left(\frac{1}{|2\Delta|}\right)^{2(N-n-1)}.
\end{eqnarray}
These equations are solved inwards from the extremal $n=0$ and $n=N$ sectors, where $\ket{\Psi_D(0)} = \ket{00\dots 00}$ and $\ket{\Psi_D(N)} = \ket{11\dots 11}$, to give
\begin{equation} 
p_n = p \left(\frac{1}{|2\Delta|}\right)^{2|n-N/2|^2}, \label{eq:p_predict}
\end{equation}
for $n=0,1,\dots, N$ and where $p = p_{N/2}$ is fixed by the normalization condition $\sum_{n=0}^N p_n = 1$. In \fir{appfig3}(a) these predicted probabilities of occupation for each sector $n$ are plotted against the exact values for the NESS with $N=6$ and $\Delta = 10$, and found to yield excellent agreement aside from the extremal sectors. Owing to the hierarchy of approximate dark states, this indicates that the NESS will be predominately a mixture of $\ket{\Psi_D(n)}$ peaked around the ``best" dark state with $n = N/2$. This result also predicts that to lowest order in $|\Delta|^{-1}$, the purity of $\rho$ is given by 
\begin{equation} \label{eq:purity}
\text{tr}(\rho^2)\approx p_{N/2}^2\approx1-\frac{1}{|\Delta|^2}+\ldots,
\end{equation}
which is independent of the size of the system. In \fir{appfig3}(b) this prediction is plotted against the exact value of $1- \textrm{tr}(\rho^2)$ of the NESS for $N=6$ as a function of $\Delta$, again showing excellent agreement even as $\Delta \rightarrow 1$. 

\begin{figure}
\includegraphics{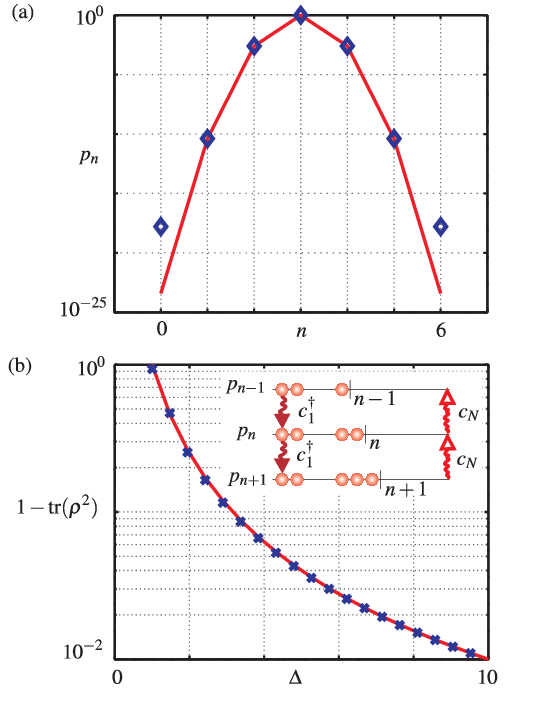}
\caption{(a) The probability $p_n$ on a logarithmic scale of the NESS $\rho$ occupying the $n$ particle sector is shown for an exact calculation ($\diamond$) for $N=6$ and $\Delta = 10$, and the predicted form (solid line) given in \eqr{eq:p_predict}. (b) The purity $1- \textrm{tr}(\rho^2)$ against $\Delta$ of the NESS $\rho$ on a logarithmic scale for an exact calculation ($\times$) for $N=6$ and the predicted form (solid line) given in \eqr{eq:purity}. The inset shows a schematic of the detailed balance approximation applied to compute these predictions.}
\label{appfig3} 
\end{figure}

\section{Equivalence of dephasing and driving effects in the toy model at large driving} \label{appendix2}

The correspondence between a slight increase of dephasing from zero and a slight decrease of the driving $f$ from unity, indicated in Eq. \eqref{eq:gamma_f}, is confirmed in \fir{appfig4}(a) in an exact calculation for the toy model with $K=20$. This equivalence is understood by considering the mean backward flow process introduced when driving slightly below $f=1$. In this limit the rate of driving from $\ket{K} \rightarrow \ket{s}$ taking the state out of $\ket{\Psi_D}$ is given by $\half(1-f)\Gamma$ and therefore slow. In contrast the rate of driving $\ket{s} \rightarrow \ket{K}$ back is $\half(1+f)\Gamma$, and therefore rapid. Focusing on the dynamics of these two driving processes alone, as in \fir{appfig4}(b), we suppose that the initial state of the system over the configurations $\ket{s}, \ket{K-1}$ and $\ket{K}$ has the form
\begin{equation}
\rho(0) = \left(\begin{array}{ccc}
0 & 0 & 0  \\
0 & \rho_{K-1}(0) & \rho_{K-1,K}(0)  \\
0 & \rho_{K-1,K}^*(0) & \rho_{K}(0) 
\end{array} \right),
\end{equation}
where there is some coherence between $\ket{K-1}$ and $\ket{K}$, but no population initially in $\ket{s}$. Evolving this state according only to these driving processes yields solutions
\begin{align} 
\begin{split}
\rho_{s}(t) &= \half(1-f)\Gamma(1 - e^{-\Gamma t})\rho_{K}(0), \\
\rho_{K}(t) &= e^{-\Gamma t}\rho_{K}(0) + \half(1+f)\Gamma(1 - e^{-\Gamma t})\rho_{K}(0),  \\
\rho_{K-1}(t) &= \rho_{K-1}(0), \\
\rho_{K-1,K}(t) &= e^{-\frac{1}{4}(1-f)\Gamma  t}\rho_{K-1,K}(0). 
\end{split}
\end{align} 
We therefore find that for the populations the stationary $t \rightarrow \infty$ limit is approached at a rate $\Gamma$ to give 
\begin{align}
\begin{split}
&\rho_{s}(\infty) = \half(1-f)\Gamma \rho_{K}(0)\\
&\rho_{K}(\infty) = \half(1+f)\Gamma \rho_{K}(0), 
\end{split}
\end{align}
while the coherence $\rho_{K-1,K}(t)$ decay exponentially to zero at a rate $\frac{1}{4}(1-f)\Gamma$. Now, by considering only the dephasing process we instead find that the populations are unchanged while the coherence decays as 
\begin{equation}
\rho_{K-1,K}(t) = e^{-2\gamma t}\rho_{K-1,K}(0). 
\end{equation}
Matching of these decoherence rates again yields \eqr{eq:gamma_f}. We therefore conclude that the emergence of a non-zero current when reducing $f$ from unity is, to leading order, caused by the resulting decoherence of the dark state $\ket{\Psi_D}$, identical to the effect of dephasing alone. This behaviour with $f$ around $f=1$, combined with $\av{J}=0$ at $f=0$ and the continuity of $\av{J}$ with $f$, is already enough to imply that NDC behaviour will be observed in the current-driving profile. Thus NDC and dephasing enhancement are underpinned by the same mechanism.

\begin{figure}[h]
\includegraphics{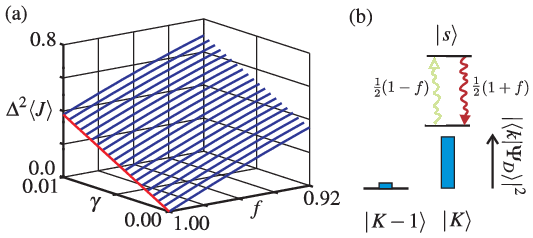}
\caption{(a) The NESS current $\av{J}$, rescaled by $\Delta^2$, for $K=20$ is shown as a function of the driving $f$ and dephasing $\gamma$ zoomed in around the limit $f=1$, $\gamma=0$ insulating point (see \fir{fig8}(a)). The behaviour at $f=1$ with $\gamma$ is emphasized by the additional (red) line. (b) In the strong driving limit $f \sim 1$, the $\Lambda_R^-$ process $\ket{K} \rightarrow \ket{s}$ occurs at a rate $\propto (1 - f) \ll 1$ making it slow (as indicated by being faded out), while the $\Lambda_R^+$ process $\ket{s} \rightarrow \ket{K}$ occurs at a rate $\propto f \approx 1$ and so is rapid. The dominant driving process is therefore $\ket{K} \rightarrow \ket{s} \rightarrow \ket{K}$, which has the effect of destroying any coherence $\ket{K}$ has with other states, such as $\ket{K-1}$.}
\label{appfig4} 
\end{figure}


\begin{thebibliography}{99}
\bibitem{footnote1} The validity of the master equation description in this regime was established in earlier works~\cite{benenti09a,benenti09b,prosen11a,znidaric11b}.
\end{thebibliography}
\end{document}